\documentclass[preprint]{iacrtrans}
\usepackage{colortbl}
\usepackage[utf8]{inputenc}
\usepackage{mathtools}
\usepackage{amsmath}
\usepackage{algorithm}
\usepackage{algorithmicx}
\usepackage{algpseudocode}
\usepackage{qcircuit}
\usepackage{listings}
\DeclarePairedDelimiter\ceil{\lceil}{\rceil}
\DeclarePairedDelimiter\floor{\lfloor}{\rfloor}

\DeclarePairedDelimiter\ket{\lvert}{\rangle}
\DeclarePairedDelimiterX\braket[2]{\langle}{\rangle}{#1 \delimsize\vert #2}

\begin{document}

\author{Hyeonhak Kim\inst{1}, Seokhie Hong\inst{2}}
\institute{Korea University, Seoul, Republic of Korea, \email{gusgkr0117@korea.ac.kr} \and 
 Korea University, Seoul, Republic of Korea, \email{shhong@korea.ac.kr}}
% \author{No Author Given}
% \institute{No Institute Given}

\title[New Space-Efficient Quantum Algorithm for Binary Elliptic Curves]{New Space-Efficient Quantum Algorithm for Binary Elliptic Curves using the Optimized Division Algorithm}

\maketitle

% use optional argument because the \LaTeX command breaks the PDF keywords
\keywords[Quantum cryptanalysis, elliptic curves, quantum resource estimation, Shor's algorithm]{Quantum cryptanalysis and elliptic curves and quantum resource estimation and Shor's algorithm}

\begin{abstract}
  In previous research, quantum resources were concretely estimated for solving Elliptic Curve Discrete Logarithm Problem(ECDLP). In \cite{banegas2020concrete}, the quantum algorithm was optimized for the binary elliptic curves and the main optimization target was the number of the logical qubits. The division algorithm was mainly optimized in \cite{banegas2020concrete} since every ancillary qubit is used in the division algorithm. In this paper, we suggest a new quantum division algorithm on the binary field which uses a smaller number of qubits. For elements in a field of $2^n$, we can save $\ceil*{n/2}-1$ qubits instead of using $8n^2+4n-12+(16n-8)\floor*{\log(n)}$ more Toffoli gates, which leads to a more space-efficient quantum algorithm for binary elliptic curves.
 
\end{abstract}

\section{Introduction}

The quantum computers currently being developed are expected to obtain the solution to a Discrete Log Problem(DLP) in a polynomial time shortly. Using Shor's algorithm\cite{365700}, we can find the solution exponentially faster than the classical algorithm. Every DLP-based cryptosystem such as RSA encryption or Diffie-Hellman key exchange is threatened by quantum computers. One of the variants of DLP is the Elliptic Curve Discrete Logarithm Problem(ECDLP) which is more useful in the cryptosystem as it needs a relatively smaller key size. Although ECC is more efficient and faster than RSA-based cryptosystems, it is more vulnerable than RSA-based cryptosystems in the quantum world. To prepare for quantum computers, we should carefully analyze the quantum security of ECC.

Although current quantum computers have a limited number of qubits and gates, which is not enough to attack currently used ECC, the day will come when quantum computers have sufficient power. Currently, the major concern in the quantum computing field is to extend the number of logical qubits. In practice, error correction must be implemented to make a logical qubit, making it harder to implement many logical qubits. According to \cite{roetteler2017quantum} and \cite{banegas2020concrete}, the security of ECC can be broken with less number of qubits than RSA-based DLP. To carefully predict the future of our cryptosystems, a concrete estimation of the quantum resource needed to solve the problem is essential.

Unlike classical circuits, quantum circuits must be reversible. It is not only to reuse the qubits but also to prevent the result of the quantum algorithm from being corrupted by the entangled qubits. Entangled qubits are involved in the algorithm even if they are not needed and not used. It's why we can't ignore or discard a part of the qubits we used. Such a quantum feature gives us the additional challenge of optimizing the quantum division algorithm.

\subsection{Contribution of this paper}

This paper proposes a space-efficient Shor's algorithm for binary elliptic curves by reducing the number of qubits in the division algorithm. We can reduce the total number of qubits by $\ceil*{n/2} - 1$ qubits for a binary elliptic curve over a field of $2^n$ elements.

In \cite{banegas2020concrete}, two distinct division algorithms are proposed. First one is based on the constant-time extended Euclidean algorithm(\cite{bernstein2019fast}). It uses fewer qubits instead of more Toffoli gates. The second one is based on Fermat's Little Theorem(FLT) and it uses fewer Toffoli gates instead of more qubits. In the quantum resource estimation, the author focuses on the space efficiency of the algorithm using the first GCD-based division algorithm. As the dominant factor of the number of qubits is the division algorithm, $7n + \floor*{\log{n}} + 9$ qubits are needed in total.

We also focus on the scale of qubits used in Shor's algorithm. As a result, using our new division algorithm, total $6n + \floor*{n/2} + \floor*{\log{n}} + 10$ qubits are needed for ECDLP, which is less than the previous algorithm in \cite{banegas2020concrete} by using $8n^2+4n-12+(16n-8)\floor*{\log(n)}$ more Toffoli gates.

\subsection{Organization of this paper}

In Section \ref{Binary elliptic curves}, we introduce the background on binary elliptic curves and Shor's algorithm to solve ECDLP. The circuit for Shor's algorithm in Section \ref{Binary elliptic curves} explains how the optimized division algorithm efficiently reduces the number of qubits. In Section \ref{Constant-time GCD algorithm}, we describe the constant-time GCD algorithm proposed in \cite{bernstein2019fast} on which our algorithm is based. Section \ref{Space-efficient Division Algorithm} suggests our new division algorithm and compares it with the original division algorithm proposed in \cite{banegas2020concrete}. Section \ref{Results} presents the quantum resource estimation result which compares with the original division algorithm. Finally, we conclude our work in Section \ref{Conclusion}.
\section{Binary elliptic curves}
\label{Binary elliptic curves}
Binary elliptic curves are elliptic curves defined over $\mathbb{F}_{2^n}$. Binary elliptic curves are given by Weierstrass form $E : y^2 + xy = x^3 + ax^2 +b$ where $a \in \mathbb{F}_2$ and $b \in \mathbb{F}_{2^n}$. The elliptic curve group of $E$ consists of the points on the curve $E$ and the point at infinity which is the identity of the group. In this section, we show a brief introduction to the arithmetic of the binary elliptic curve group and Shor's algorithm for ECDLP.

\subsection{Arithmetic of binary elliptic curves}
A point on the elliptic curve $E$ is a tuple $(x,y)$ where $x, y \in \mathbb{F}_{2^n}$. The arithmetic of the field $\mathbb{F}_{2^n}$ is used to compute the point addition on the curve. A finite field $\mathbb{F}_{2^n}$ is given by an irreducible polynomial $m(x) \in \mathbb{F}_2[x]$ of degree $n$. Several standard irreducible polynomials for each security level(163, 233, 283, 409, 571) are specified in \cite{FIPS186-4}.

For a point $P_1 = (x_1, y_1) \in E$, the inverse of the point $P_1$ is $-P_1 = (x_1, y_1 + x_1)$. For the points $P_1 = (x_1, y_1)$, $P_2 = (x_2, y_2) \neq \pm P_1$ and $P_3 = (x_3, y_3) = P_1 + P_2$, the addition is computed as follows.
$$x_3 = \lambda^2 + \lambda + x_1 + x_2 + a,\:y_3 = (x_2 + x_3)\lambda + x_3 + y_2$$
where $\lambda = (y_1+y_2)/(x_1 + x_2)$. We define the point at infinity $O$ as $O = P_1 + (-P_1)$ which is the identity of the elliptic curve group. For the doubling operation $P_3 = [2]P_1$, we use the slightly different formula below.
$$x_3 = \lambda^2 + \lambda + a,\:y_3 = x_1^2 + (\lambda + 1)x_3$$
where $\lambda = x_1 + y_1/x_1$.

\subsection{Shor's algorithm for ECDLP}
Hidden subgroup problems for finite Abelian groups such as integer factorization and discrete logarithm can be solved by Shor's algorithm in polynomial time. In the elliptic curve cryptography, we need to find the solution $r\in\mathbb{Z}$ such that $[r]P = Q$ for two distinct points $P, Q \in E$.
Shor's algorithm uses quantum Fourier transform preceding a final measurement. We can optimize the scale of the qubits using semiclassical Fourier transform in \cite{griffiths1996semiclassical} as figure \ref{Semiclassical quantum circuit}.

\begin{figure}[H]
\centerline{
\Qcircuit @C=0.3em @R=0.7em @!R {
 & & & & \ustick{\mu_{0}} \cwx[1] & & & & & & \ustick{\mu_{1}} \cwx[1] & & & & & & & \ustick{\mu_{2n+1}} \cwx[1] \\ 
\lstick{\ket{0}} & \gate{H} & \ctrl{1} & \gate{H} & \meter & \push{\ket{0}} & \gate{H} & \ctrl{1} & \gate{R_{\theta_{1}}} & \gate{H} & \meter & \push{\cdots} & \push{\ket{0}}& \gate{H} & \ctrl{1} & \gate{R_{\theta_{2n+1}}} & \gate{H} & \meter \\
\lstick{\ket{O}} & {/^{2n}} \qw & \gate{+P} & \qw & \qw & \qw & \qw & \gate{+[2]P} & \qw & \qw & \qw & \push{\cdots} &  & \qw & \gate{+[2^n]Q} & \qw & \qw & \qw \\
}
}
\caption{Shor's algorithm for ECDLP witha semiclassical Fourier transform}
\label{Semiclassical quantum circuit}
\end{figure}

Unlike the usual quantum Fourier transform which needs $2n+2$ zero-initialized qubits, the semiclassical Fourier transform only needs one zero-initialized qubit. We can see that it is important to implement an efficient constant point addition gate to optimize Shor's algorithm. The quantum circuit of the constant point addition for binary elliptic curves was constructed in \cite{banegas2020concrete} as figure \ref{Constant point addition circuit} below. In figure \ref{Constant point addition circuit}, addition, multiplication, and squaring gates don't require any ancillary qubits. Only the modulo division gates need ancillary qubits. That's the reason why we optimize the modulo division gate to reduce the number of ancillary qubits required.

\begin{figure}[H]
\centerline{
\Qcircuit @C=0.5em @R=0.7em @!R {
\lstick{\ket{x_1}} & {/^{n}} \qw & \gate{+x_2} & \ctrl{3} & \ctrl{2} & \gate{+a+x_2} & \targ & \targ & \qw & \ctrl{2} & \ctrl{3} & \gate{+x_2} & \ctrl{2} & \rstick{\ket{x_3} or \ket{x_1}} \qw \\
\lstick{\ket{q}} & \qw & \ctrl{1} & \qw & \qw & \ctrl{-1} & \ctrl{-1} & \ctrl{-1} & \qw & \qw & \qw & \ctrl{1} & \ctrl{1} & \rstick{\ket{q}} \qw \\
\lstick{\ket{y_1}} & {/^{n}} \qw & \gate{y_2} & \ctrl{1} & \gate{M} & \gate{S} & \ctrl{-2} & \qw & \gate{S} & \gate{M} & \ctrl{1} & \gate{+y_2} & \targ & \rstick{\ket{y_3} \textrm{or} \ket{y_1}} \qw \\
\lstick{\ket{0}} & {/^{n}} \qw & \qw & \gate{D} & \ctrl{-1} & \ctrl{-1} & \qw & \ctrl{-3} & \ctrl{-1} & \ctrl{-1} & \gate{D} & \qw & \qw & \rstick{\ket{0}} \qw\\
}
}
\caption{Constant point addition circuit. $M$ is MODMULT, $S$ is squaring with separate output, $D$ is division.\cite{banegas2020concrete}}
\label{Constant point addition circuit}
\end{figure}

\section{Constant-time GCD algorithm}
\label{Constant-time GCD algorithm}
The original Euclid GCD algorithm for binary field $\mathbb{F}_{2^n}$ is not a constant-time algorithm as the algorithm checks if the leading coefficient is zero or not. When we design a quantum circuit, implementing conditional branches becomes more daunting. It generally needs more ancillary qubits and Toffoli gates. Thus constant-time GCD algorithm uses fewer qubits. As in \cite{banegas2020concrete}, we use constant-time GCD algorithm in \cite{bernstein2019fast} to reduce the number of ancillary qubits. In this section, we briefly demonstrate the constant-time GCD algorithm and the constant-time modulus inversion algorithm.

\subsection{Iterates of Division steps}

Define $\mathrm{divstep} : \mathbb{Z} \times k[[x]]^* \times k[[x]] \rightarrow \mathbb{Z} \times k[[x]]^* \times k[[x]]$ as follows:

\begin{equation}
  \mathrm{divstep}(\delta, f, g) =
    \begin{cases}
      (1-\delta, g, (g(0)f - f(0)g)/x) & \text{if $\delta > 0$ and $g(0) \neq 0$}\\
      (1+\delta, f, (f(0)g - g(0)f)/x) & \text{otherwise}\\
    \end{cases}       
\end{equation}

For $\mathrm{divstep}$, there are corresponding transition matrices. Suppose $(\delta_n, f_n, g_n) = \mathrm{divstep}(\delta, f, g)$. Then the transition matrices are
\begin{equation}
    \begin{pmatrix}
        f_{n+1}\\ 
        g_{n+1}
    \end{pmatrix}=
    \mathcal{T}(\delta_{n}, f_{n}, g_{n})
    \begin{pmatrix}
        f_{n}\\
        g_{n}
    \end{pmatrix} \quad \textrm{and} \quad 
    \begin{pmatrix}
        1\\ 
        \delta_{n+1}
    \end{pmatrix}=
    \mathcal{S}(\delta_{n}, f_{n}, g_{n})
    \begin{pmatrix}
        1\\
        \delta_{n}
    \end{pmatrix}
\end{equation}

where $\mathcal{T} : \mathbb{Z} \times k[[x]]^* \times k[[x]] \rightarrow M_2(k[1/x])$ is defined by

\begin{equation}
    \mathcal{T}(\delta, f, g)=
    \begin{dcases}
        \begin{pmatrix}
        0 & 1\\
        \frac{g(0)}{x} & \frac{-f(0)}{x}\\
        \end{pmatrix} & \text{if $\delta > 0$ and $g(0) \neq 0$}\\
        \begin{pmatrix}
        1 & 0\\
        \frac{-g(0)}{x} & \frac{f(0)}{x}\\
        \end{pmatrix} & \text{otherwise}\\
    \end{dcases}
\end{equation}

and $\mathcal{S} : \mathbb{Z} \times k[[x]]^* \times k[[x]] \rightarrow M_2(\mathbb{Z})$ is defined by

\begin{equation}
    \mathcal{S}(\delta, f, g)=
    \begin{dcases}
        \begin{pmatrix}
        1 & 0\\
        1 & -1\\
        \end{pmatrix} & \text{if $\delta > 0$ and $g(0) \neq 0$}\\
        \begin{pmatrix}
        1 & 0\\
        1 & 1\\
        \end{pmatrix} & \text{otherwise}\\
    \end{dcases}
\end{equation}

In order to find a greatest common divisor of two polynomials $R_0, R_1$, we will use the reversed polynomials $f = x^{d}R_0(1/x)$ and $g = x^{d-1}R_1(1/x)$ where $d = \deg{R_0} > \deg{R_1}$. Using the reversed polynomials $f$ and $g$, we can find the greatest common divisor of two polynomials $R_0,R_1$ by repeating the divstep. We can also find the inverse modulo $R_0$ of the given polynomial $R_1$ accordingly. The theorem \ref{divstep theorem} below shows the relationship between the output of the divstep algorithm and the desired greatest common divisor.

\begin{theorem}[Theorem 6.2 in \cite{bernstein2019fast}]
\label{divstep theorem}
Let \(k\) be a field. Let \(d\) be a positive integer. Let $R_0, R_1$ be elements of the polynomial ring $k[x]$ with $\deg(R_0) = d > \deg{R_1}$. Define $G=\gcd(R_0, R_1)$, and let $V$ be the unique polynomial of degree $< d - \deg{G}$ such that $VR_1 \equiv G (\mod{R_0})$. Define $f = x^{d}R_0(1/x); g = x^{d-1}R_1(1/x); (\delta_n, f_n, g_n) = \mathrm{divstep}^{n}(1, f, g); \mathcal{T}_n = \mathcal{T}(\delta_n, f_n, g_n);$ and  $\begin{pmatrix} u_n & v_n\\ q_n & r_n\end{pmatrix}=\mathcal{T}_{n-1}...\mathcal{T}_{0}$. Then

\begin{equation}
\begin{split}
    \deg{G} &= \delta_{2d-1}/2\\
    G &= x^{\deg{G}}f_{2d-1}(1/x)/f_{2d-1}(0)\\
    V &= x^{-d+1+\deg{G}}v_{2d-1}(1/x)/f_{2d-1}(0)
\end{split}
\end{equation}
\end{theorem}

We note that $\begin{pmatrix}v_n\\ r_n\end{pmatrix} = \mathcal{T}(\delta_{n-1}, f_{n-1}, g_{n-1})\begin{pmatrix}v_{n-1}\\ r_{n-1}\end{pmatrix}$ where $\begin{pmatrix}v_0\\ r_0\end{pmatrix}=\begin{pmatrix}0\\ 1\end{pmatrix}$ and using this recurrence relation, we can find the modulus inverse $V$ by computing $V = x^{-d+1+\deg{G}}v_{2d-1}(1/x)/f_{2d-1}(0)$. We apply theorem \ref{divstep theorem} to binary field and show how to construct an efficient modulus inversion algorithm for binary field on the next subsection.

\subsection{Modulus Inversion Algorithm in Binary Field using Division steps}
Suppose we have an irreducible polynomial $R_0(x) \in \mathbb{F}_2[x]$ with degree $d$. We will find the inverse modulo $R_0(x)$ of a non-zero polynomial $R_1(x) \in \mathbb{F}_2[x]$ with $\deg{R_1} < d$. As we know that $G = \gcd(R_0, R_1) = 1$, we can compute the modulus inversion $V = x^{-d+1}v_{2d-1}(1/x)$ by theorem \ref{divstep theorem}. We will use the three equations below to compute the $v_{2d-1}$.

\begin{equation}
\label{eq:6}
\begin{split}
    \begin{pmatrix}v_n\\ r_n\end{pmatrix} &= \mathcal{T}(\delta_{n-1}, f_{n-1}, g_{n-1})\begin{pmatrix}v_{n-1}\\ r_{n-1}\end{pmatrix}\\
    \begin{pmatrix}f_{n}\\g_{n}\end{pmatrix} &= \mathcal{T}(\delta_{n-1}, f_{n-1}, g_{n-1})
    \begin{pmatrix}f_{n-1}\\g_{n-1}\end{pmatrix}\\
    \begin{pmatrix}1\\\delta_{n}\end{pmatrix} &= \mathcal{S}(\delta_{n-1}, f_{n-1}, g_{n-1})
    \begin{pmatrix}1\\\delta_{n-1}\end{pmatrix}
\end{split}
\end{equation}

where $v_0 = 1, r_0 = 1, f_0 = f, g_0 = g, \delta_0 = 1$. Using the three recurrence relation \ref{eq:6}, we can make a constant-time modulus inversion algorithm in binary field. The figure \ref{fig:modinverse algorithm} below illustrates the algorithm. The algorithm is expressed in the Sage\cite{sagemath} computer-algebra system.

\begin{figure}[H]
\label{fig:modinverse algorithm}
\caption{The algorithm modinverse using $2d-1$ divsteps in binary field}
\begin{lstlisting}[language=Python]
def modinverse(R0,R1):
  d = R0.degree()
  assert d > 0 and d > R1.degree() and R1 != 0
  assert R0 is irreducible
  f,g = R0.reverse(d), R1.reverse(d-1)
  kx = f.parent()
  x = kx.gen()
  delta,v,r = 1,kx(0),kx(1)
    
  n = 2*d-1
  while n > 0:
    if delta > 0 and g[0] == 1: delta,f,g,v,r = -delta,g,f,r,v
    f0,g0 = f[0],g[0]
    delta,g,r = 1+delta,(g + g0*f)/x,(r + g0*v)/x
    n = n-1
    
  return kx(x^(2*d-2)*v).reverse(d-1)
\end{lstlisting}
\end{figure}

The modinverse algorithm consists of swap, addition, and shift operations in the binary extension field, which makes it easy for us to make a circuit for the algorithm. Although the algorithm can be converted into a classical circuit, the algorithm needs to be reversible in order to make a corresponding quantum circuit. In the next section, we demonstrate how the previous quantum algorithm was constructed in \cite{banegas2020concrete} and how we optimize it in a space-efficient way.
\section{Space-efficient Division Algorithm}
\label{Space-efficient Division Algorithm}
The division algorithm is the bottleneck of Shor's algorithm. Every ancillary qubit is used in the division step. There are two different ways to compute an inverse in a binary field as suggested in \cite{banegas2020concrete}: GCD\_DIV, FLT\_DIV. In this paper, we focus on the GCD\_DIV algorithm to improve space efficiency.

The GCD\_DIV algorithm needs a quantum gate for incrementing an integer. In \cite{banegas2020concrete}, the author used the increment gate described in \cite{gidney2015constructing}. This gate needs $n$ borrowed qubits which are the ancillary qubits that are not necessarily initialized, which saves the number of qubits. The increment gate uses $4n-4$ TOF gates and $10n-6$ CNOT gates. As this can save the number of qubits significantly, we used the same way in our new algorithm. We also used the large controlled-NOT gate($C^n\mathrm{NOT}$) described in \cite{gidney2015constructing} which needs $n-2$ borrowed qubits for a $C^n\mathrm{NOT}$ gate. This $C^n\mathrm{NOT}$ gate uses $3n-6$ Toffoli gates.

\subsection{Previous Division Algorithm}
For computing division in a field $\mathbb{F}_{2^n}$, the algorithm in \cite{banegas2020concrete} uses the constant-time extended GCD algorithm introduced in \cite{bernstein2019fast}. This algorithm uses $4n + \floor*{\log{n}} + 8$ ancillary qubits plus $3n$ qubits for the input and output qubits. Not including the multiplication step(line 25), Algorithm \ref{PreviousGCDAlgorithm} uses $12n^2 + 116n - 62 + (88n - 44)\floor*{\log(n)}$ Toffoli gates.

\begin{algorithm}[H]
\caption{GCD\_DIV\cite{banegas2020concrete}}\label{PreviousGCDAlgorithm}

\textbf{Fixed input : } A constant field polynomial $m$ of degree $n > 0$ as an array $M$.\\
                        $\Lambda = min(2n-2-\ell, n)$, $\lambda = min(\ell+1 , n)$.
                        
\textbf{Quantum input : }
\begin{itemize}
    \setlength\itemsep{-0.4em}
    \item A non-zero binary polynomial $R_1(z)$ of degree up to $n-1$ stored in array $g$ of size $n$ to invert.
    \item A binary polynomial $R_2(z)$ of degree up to $n-1$ to multiply with the inverse stored in array $B$.
    \item A binary polynomial $R_3(z)$ of degree up to $n-1$ for the result stored in array $C$.
    \item 4 arrays of size $n+1$ : $f, r, v, g_0$ initialized to an all-$\ket{0}$ state.
    \item 1 array of size $\floor*{\log{n}} + 2$ initialized to an all-$\ket{0}$ state : $\delta$, which will be treated as an integer
    \item 2 qubits to store ancillary qubits $a, g[n]$ initialized to $\ket{0}$.
    \item Refer to $g[n], g[n-1], ..., g[3]$ as $g_0[n+1], g_0[n+2], ...,g_0[2n-2]$ when applicable.
    \item Refer to $\delta[\floor*{\log{n}} + 1]$ as sign with sign$=1$ if $\delta > 2^{\floor*{\log{n}}+1}$ and 0 otherwise.
\end{itemize}

\textbf{Result : } Everything except $C$ the same as their input, $C$ as $R_3 + R_2/R_1$

\begin{algorithmic}[1]
\For{$i$ in $M$}
\State{$f[n-i] \leftarrow \ket{1}$}
\EndFor
\State{$\mathrm{sign} \leftarrow \ket{1}$}
\State{$r[0] \leftarrow \ket{1}$}
\For{$i=0,...,\floor*{\frac{n}{2}}-1$}
\State{$\mathrm{SWAP}(g[i], g[n-1-i])$}
\EndFor
\For{$\ell=0,...,2n-2$}
\State{$v[0,...,n] \leftarrow \mathrm{RIGHTSHIFT}(v[0,...,n])$}
\State{$a \leftarrow \mathrm{TOF}(\mathrm{sign}, g[0], a)$}
\State{$\delta[0,...,\floor*{\log{n}}+1] \leftarrow \mathrm{CNOT}(\delta[0,...,\floor*{\log{n}}+1], a)$}
\State{$\mathrm{CSWAP}_a(f[0,...,\Lambda], g[0,...,\Lambda])$}
\State{$\mathrm{CSWAP}_a(r[0,...,\lambda], v[0,...,\lambda])$}
\State{$\delta[0,...,\floor*{\log{n}}+1] \leftarrow \mathrm{INC}_{1+a}(\delta[0,...,\floor*{\log{n}}+1])$}
\State{$a \leftarrow \mathrm{CNOT}(a, v[0])$}
\State{$g_0[\ell] \leftarrow \mathrm{CNOT}(g_0[\ell], g[0])$}
\algstore{alg1}
\end{algorithmic}
\end{algorithm}

\begin{algorithm}
\begin{algorithmic}[1]
\algrestore{alg1}
\State{$g[0,...,\Lambda] \leftarrow \mathrm{TOF}(f[0,...,\Lambda],g_0[\ell],g[0,...,\Lambda])$}
\State{$r[0,...,\lambda] \leftarrow \mathrm{TOF}(v[0,...,\lambda],g_0[\ell],r[0,...,\lambda])$}
\State{$g[0,...,\Lambda] \leftarrow \mathrm{LEFTSHIFT}(g[0,...,\Lambda])$}
\EndFor
\For{$i=0,...,\floor*{\frac{n}{2}}-1$}
\State{$\mathrm{SWAP}(v[i], v[n-1-i])$}
\EndFor
\State{$C[0,...,n-1] \leftarrow \mathrm{MODMULT}(v[0,...,n-1], B[0,...,n-1], C[0,...,n-1])$}
\State{UNCOMPUTE lines 1-24}
\end{algorithmic}
\end{algorithm}

\begin{figure}[H]
\centerline{
\Qcircuit @C=2em @R=1em @!R {
\lstick{\ket{\delta}} & {/^{|\delta|}} \qw & \qw & \targ & \qw & \qw & \multigate{1}{+1} & \qw & \qw & \qw & \rstick{\ket{\delta}} \qw \\
\lstick{\ket{\mathrm{sign}}} & \qw & \ctrl{4} & \targ & \qw & \qw & \ghost{+1} & \qw & \qw & \qw & \rstick{\ket{\mathrm{sign}}} \qw \\
\lstick{\ket{f}} & {/^{n+1}} \qw & \qw & \qw & \qswap & \qw & \qw & \qw & \ctrl{1} & \qw & \rstick{\ket{f}} \qw \\
\lstick{\ket{g}} & {/^{n+1}} \qw & \ctrlo{2} & \qw & \qswap \qwx & \qw & \qw & \ctrl{1} & \targ & \gate{<} & \rstick{\ket{g}} \qw \\
\lstick{g_0[\ell] = \ket{0}} & \qw & \qw & \qw & \qw & \qw & \qw & \targ & \ctrl{-1} & \ctrl{2} & \rstick{\ket{g_0[\ell]}} \qw \\
\lstick{a = \ket{0}} & \qw & \targ & \ctrl{-5} & \ctrl{-2} & \ctrl{1} & \ctrlo{-4} & \targ & \qw & \qw & \rstick{\ket{0}} \qw \\
\lstick{\ket{r}} & {/^{n+1}} \qw & \qw & \qw & \qw & \qswap & \qw & \qw & \qw & \targ & \rstick{\ket{r}} \qw \\
\lstick{\ket{v}} & {/^{n+1}} \qw & \gate{>} & \qw & \qw & \qswap \qwx & \qw & \ctrl{-2} & \qw & \ctrl{-1} & \rstick{\ket{v}} \qw \\
}
}
\caption{A single step of Algorithm \ref{PreviousGCDAlgorithm}. $|\delta| = \floor*{\log{n}} + 2$(\cite{banegas2020concrete})}
\label{PreviousGCDCircuit}
\end{figure}

\subsection{New Division Algorithm}
Algorithm \ref{PreviousGCDAlgorithm} must use $n+1$ ancillary qubits to store $g_0[\ell]$ at each step $\ell$. As $g_0[\ell]$ is used as a control bit of each step, it cannot be discarded from the algorithm. We have to carry these qubits even if they won't be used in the future step.

We suggest algorithm \ref{NewGCDAlgorithm} to avoid using qubit register $g_0[0,...,n]$. In the new algorithm, we use the LEFTROTATE operation for $g$ instead of the LEFTSHIFT operation at line 20 in the algorithm \ref{PreviousGCDAlgorithm}. Using the LEFTROTATE operation doesn't need to consider handling the least significant bit of the input register and doesn't need $g_0[\ell]$ register to store it. Instead, we use $mask$ register for masking $f$ and $g$ registers in order to run the addition operation between $f$ and $g$ normally. The $mask$ register is only used when $f$ and $g$ are added, which means that computing a mask for $\min(f,g)$ is enough to run the algorithm. In this situation, the fact that $\deg(\min(f,g)) \leq n - 1 - \floor*{\ell/2}$ for each step $\ell$ gives us some free space to save other qubits.

The $mask$ register must be an array of size equal to or greater than $n - 1$. We can assure that the $mask[n-1,...,n-\floor*{\ell/2}]$ is all-$\ket{0}$ state at each step $\ell$. It means that the array size of the $mask$ register decreases by at least 1 qubit every two steps. On the other hand, the array size for the $\max(r,v)$ increases by at most 1 qubit per step. As we got both constantly decreasing and increasing arrays, we can partially overlap these two arrays. As a result, the algorithm \ref{NewGCDAlgorithm} uses $\ceil*{n/2} - 1$ fewer qubits than the algorithm \ref{PreviousGCDAlgorithm}.

The algorithm \ref{NewGCDAlgorithm} uses the algorithm \ref{ControlledLeftshiftAlgorithm} as a subroutine. The algorithm \ref{ControlledLeftshiftAlgorithm} is identical to the controlled left-shift algorithm except that it is only used for arrays of qubits where the first bit is 0. Figure \ref{NewGCDCircuit} shows the corresponding quantum circuit for a single step of algorithm \ref{NewGCDAlgorithm}. The circuit is also repeated $2n-1$ times.

\begin{algorithm}[H]
\caption{$\mathrm{MASK\_LEFTSHIFT}_q$}\label{ControlledLeftshiftAlgorithm}
\textbf{Fixed input : } The length $\Lambda + 1$ of the input quantum register $A$.

\textbf{Quantum input : }
\begin{itemize}
    \setlength\itemsep{-0.4em}
    \item An array $A[0,...,\Lambda]$ initialized to an all-$\ket{1}$ state.
    \item 1 qubit $\ket{q}$ for the control qubit.
\end{itemize}
\textbf{Result : } An left-shifted array $A[0,...,\Lambda]$ if $q=1$ and $\Lambda \geq 1$, the same as its input otherwise. 

\begin{algorithmic}[1]
\If{$\Lambda < 1$}
\State{$A[0,...,\Lambda] \leftarrow A[0,...,\Lambda]$}
\Else
\State{$A[0] \leftarrow \mathrm{TOF}(A[1], q, A[0])$}
\State{$A[0,...,\Lambda] \leftarrow \mathrm{LEFTROTATE}_q(A[0,...,\Lambda])$}
\EndIf
\end{algorithmic}
\end{algorithm}

\begin{algorithm}[H]
\caption{NEW GCD\_DIV}\label{NewGCDAlgorithm}

\textbf{Fixed input : } A constant field polynomial $m$ of degree $n > 0$ as an array $M$.\\
                        $\Lambda = n-1-\max(\floor*{\frac{\ell - 1}{2}}, 0), \lambda = \min(\ell, n)$.
                        
\textbf{Quantum input : }
\begin{itemize}
    \setlength\itemsep{-0.4em}
    \item A non-zero binary polynomial $R_1(z)$ of degree up to $n-1$ stored in array $g$ of size $n$ to invert.
    \item A binary polynomial $R_2(z)$ of degree up to $n-1$ to multiply with the inverse stored in array $B$.
    \item A binary polynomial $R_3(z)$ of degree up to $n-1$ for the result stored in array $C$.
    \item 2 arrays of size $n+1$ : $f, v$ initialized to an all-$\ket{0}$ state.
    \item 1 array of size $\floor*{\log{n}} + 2$ initialized to an all-$\ket{0}$ state : $\delta$, which will be treated as an integer
    \item 1 array of size $n + 2 + \floor*{n/2}$ : $mask$ initialized to an all-$\ket{0}$ state.
    \item 3 qubits to store ancillary qubits $a, b, g[n]$ initialized to $\ket{0}$.
    \item Refer to $\delta[\floor*{\log{n}} + 1]$ as sign with sign$=1$ if $\delta > 2^{\floor*{\log{n}}+1}$ and 0 otherwise.
    \item Refer to $r[0],r[1],...,r[n]$ as $mask[n + \floor*{n/2} + 1],mask[n + \floor*{n/2}],...,mask[\floor*{n/2} + 1]$.
\end{itemize}

\textbf{Result : } Everything except $C$ the same as their input, $C$ as $R_3 + R_2/R_1$

\begin{algorithmic}[1]
\For{$i$ in $M$}
\State{$f[n-i] \leftarrow \ket{1}$}
\EndFor
\For{$i=0,...,n-1$}
\State{$mask[i] \leftarrow \ket{1}$}
\EndFor
\State{$\mathrm{sign} \leftarrow \ket{1}$}
\State{$r[0] \leftarrow \ket{1}$}
\For{$i=0,...,\floor*{\frac{n}{2}}-1$}
\State{$\mathrm{SWAP}(g[i], g[n-1-i])$}
\EndFor
\For{$\ell=0,...,2n-2$}
\State{$v[0,...,n] \leftarrow \mathrm{RIGHTSHIFT}(v[0,...,n])$}
\State{$a \leftarrow \mathrm{TOF}(\mathrm{sign}, g[0], a)$}
\State{$b \leftarrow \mathrm{TOF}(\mathrm{sign}, g[0]+1, b)$}
\algstore{alg2}
\end{algorithmic}
\end{algorithm}

\begin{algorithm}
\begin{algorithmic}[1]
\algrestore{alg2}
\State{$\mathrm{CSWAP}_a(f[0,...,n], g[0,...,n])$}
\State{$\mathrm{CSWAP}_a(r[0,...,\lambda], v[0,...,\lambda])$}
\State{$g[1,...,\Lambda] \leftarrow \mathrm{CTOF}(g[0],f[1,...,\Lambda],mask[1,...,\Lambda],g[1,...,\Lambda])$}
\State{$mask[0,...,\Lambda] \leftarrow \mathrm{MASK\_LEFTSHIFT}_b(mask[0,...,\Lambda])$}
\State{$b \leftarrow \mathrm{TOF}(\mathrm{sign}, g[0]+1,b)$}
\State{$mask[0,...,\Lambda] \leftarrow \mathrm{C^nMASK\_LEFTSHIFT}_{\delta}(mask[0,...,\Lambda])$}
\State{$\delta[0,...,\floor*{\log{n}}+1] \leftarrow \mathrm{CNOT}(\delta[0,...,\floor*{\log{n}}+1], a)$}
\State{$\delta[0,...,\floor*{\log{n}}+1] \leftarrow \mathrm{INC}_{1+a}(\delta[0,...,\floor*{\log{n}}+1])$}
\State{$a \leftarrow \mathrm{TOF}(a, v[0], g[0])$}
\State{$r[0,...,\lambda] \leftarrow \mathrm{TOF}(v[0,...,\lambda],g[0],r[0,...,\lambda])$}
\State{$g[0,...,\Lambda] \leftarrow \mathrm{LEFTROTATE}(g[0,...,\Lambda])$}
\EndFor
\For{$i=0,...,\floor*{\frac{n}{2}}-1$}
\State{$\mathrm{SWAP}(v[i], v[n-1-i])$}
\EndFor
\State{$C[0,...,n-1] \leftarrow \mathrm{MODMULT}(v[0,...,n-1], B[0,...,n-1], C[0,...,n-1])$}
\State{UNCOMPUTE lines 1-30}
\end{algorithmic}
\end{algorithm}
\

\begin{remark}
Note that the $mask$ register can be LEFTSHIFT-ed using the algorithm \ref{ControlledLeftshiftAlgorithm}. Until $\deg(\min(f,g)) \geq 2$, the fact that $mask[0]$ and $mask[1]$ are $\ket{1}$ is guaranteed. So, we can eliminate the $mask[0]$ by XOR-ing with $mask[1]$ until $\deg(\min(f,g)) \geq 2$.
\end{remark}

Algorithm \ref{NewGCDAlgorithm} uses the similar strategy with algorithm \ref{PreviousGCDAlgorithm} but uses $mask$ register. The bit length of $mask$ register must decreases as the loop progresses. We shift the $mask$ register when the condition $g[0]=0$ and $\delta < 0$ or $\delta = 0$ is satisfied. The condition $g[0]=0$ and $\delta < 0$ can be verified with $b$ register and the condition $\delta = 0$ can be verified with $\delta$ register. As $\delta$ is not a single qubit, we need to use the large controlled not gate\cite{gidney2015constructing}. Thus the $\mathrm{C^nMASK\_LEFTSHIFT}$ gate uses $4\floor*{\log(n)} + \Lambda -2$ Toffoli gates, $2\Lambda$ CNOT gates and $\floor*{\log(n)}$ borrowed qubits. The quantum circuit is detailed in Appendix \ref{appendix:Large Controlled Leftshift}.

Algorithm \ref{NewGCDAlgorithm} needs controlled Toffoli gate($\mathrm{CTOF}$) at line 18 which is equivalent with $\mathrm{C^3NOT}$ gate. The controlled Toffoli gate can be implemented using 3 Toffoli gates and 1 ancillary qubit. So it needs $3n$ Toffoli gates and $n$ ancillary qubit to implement $n$ bit controlled Toffoli gate with 1 depth. In order to use minimum qubits, we can use only one ancillary qubit if we use $n$ depth. In this case, we can use $v[0]$ as an ancillary qubit before doing line 17. On the other hand, we can also apply the $\mathrm{C^3NOT}$ gate in \cite{gidney2015constructing} without any ancillary qubit if we use $4n$ Toffoli gates within $4$ depth. In both cases, we don't need additional ancillary qubits.

For any other gates such as $\mathrm{INC}_{1+a}$ in line 23 and $\mathrm{MODMULT}$ in line 31 of Algorithm \ref{NewGCDAlgorithm}, we adapt the same gates used in \cite{banegas2020concrete}. These gates don't require ancillary qubits.

\begin{figure}[ht]
\centerline{
\Qcircuit @C=1.5em @R=0.7em @!R {
\lstick{\ket{\delta}} & {/^{|\delta|}} \qw & \qw & \qw & \qw & \qw & \qw & \qw & \ctrl{5} & \targ & \multigate{1}{+1} & \qw & \qw & \qw & \rstick{\ket{\delta}} \qw \\
\lstick{\ket{\mathrm{sign}}} & \qw & \ctrl{4} & \ctrl{5} & \qw & \qw & \qw & \ctrl{5} & \qw & \qw & \ghost{+1} & \qw & \qw & \qw & \rstick{\ket{\mathrm{sign}}} \qw \\
\lstick{\ket{f}} & {/^{n+1}} \qw & \qw & \qw & \qswap & \ctrl{1} & \qw & \qw & \qw & \qw & \qw & \qw & \qw & \qw & \rstick{\ket{f}} \qw \\
\lstick{\ket{g}} & {/^{n+1}} \qw & \qw & \qw & \qswap \qwx & \targ & \qw & \qw & \qw & \qw & \qw & \qw & \qw & \multigate{1}{<<<} & \rstick{\ket{g}} \qw \\
\lstick{\ket{g[0]}} &  \qw & \ctrl{2} & \ctrlo{3} & \qw & \ctrl{-1} & \qw & \ctrlo{3} & \qw & \qw & \qw & \ctrl{2} & \ctrl{4} & \ghost{<<<} & \rstick{\ket{g[0]}} \qw \\
\lstick{\ket{mask}} & {/^{|mask|}} \qw & \qw & \qw & \qw & \ctrl{-2} & \gate{<} & \qw & \gate{<} & \qw & \qw & \qw & \qw & \qw & \rstick{\ket{mask}} \qw \\
\lstick{a = \ket{0}} & \qw & \targ & \qw & \ctrl{-3} & \ctrl{2} & \qw & \qw & \qw & \ctrl{-6} & \ctrlo{-5} & \targ & \qw & \qw & \rstick{\ket{0}} \qw \\
\lstick{b = \ket{0}} & \qw & \qw & \targ & \qw & \qw & \ctrl{-2} & \targ & \qw & \qw & \qw & \qw & \qw & \qw & \rstick{\ket{0}} \qw \\
\lstick{\ket{r}} & {/^{n+1}} \qw & \qw & \qw & \qw & \qswap & \qw & \qw & \qw & \qw & \qw & \qw & \targ & \qw & \rstick{\ket{r}} \qw \\
\lstick{\ket{v}} & {/^{n+1}} \qw & \gate{>} & \qw & \qw & \qswap \qwx & \qw & \qw & \qw & \qw & \qw & \ctrl{-3} & \ctrl{-1} & \qw & \rstick{\ket{v}} \qw \\
}
}
\caption{A single step of Algorithm \ref{NewGCDAlgorithm}. $|\delta| = \floor*{\log{n}} + 2$ and $|mask| = n + 2 + \floor*{n/2}$}
\label{NewGCDCircuit}
\end{figure}

As a result, algorithm \ref{NewGCDAlgorithm} uses $3n + \floor*{n/2} + \floor*{\log(n)} + 9$ ancillary qubits and $3n$ qubits for input and output, which is $n - \floor*{n/2} - 1 = \ceil*{n/2}-1$ fewer than the algorithm \ref{PreviousGCDAlgorithm}. Also, not including the multiplication step(line 31), algorithm \ref{NewGCDAlgorithm} uses $20n^2 + 120n - 74 + (104n-52)\floor*{\log(n)}$ Toffoli gates\footnote{We used SageMath to verify the number of Toffoli gates and CNOT gates. Our code is available at https://github.com/gusgkr0117/space-efficient-quantum-division-algorithm-in-binary-field}, which is $8n^2+4n-12+(16n-8)\floor*{\log(n)}$ more than the algorithm \ref{PreviousGCDAlgorithm}. We can see that the new algorithm is more space-efficient than the previous algorithm even though it uses more Toffoli gates.

The main feature of the new algorithm is the fact that the array of $mask$ and the array of $r$ share $\ceil*{n/2} - 1$ qubits. To ensure that the algorithm is well-designed, we can see the table \ref{GCD_DIV example} as an example. The table \ref{GCD_DIV example} shows the state of the quantum registers at each step $\ell$.

For example, if we want to find an inverse of $g$ where $f=x^8 + x^4 + x^2 + x^1 + 1$, $g=x^7 + x^4 + 1$, the state of the quantum registers at each step are described in table \ref{GCD_DIV example}. We can see that the $m_1 \leq \Lambda$ and $m_2 \leq \lambda$ for every step $0 \leq l \leq 2n-1$, which means that the value of $v$ and the value of $mask$ do not interfere with each other in the overlapped array in the whole steps.

\begin{table}[H]
{
\centerline{
\caption{Iterates $(\delta, f, g, m_1, \Lambda, m_2, \lambda)$ for $n=8, f=x^8 + x^4 + x^2 + x^1 + 1$, $g=x^7 + x^4 + 1$ where $m_1 = \min(\deg(f),\deg(g))$, $\Lambda = n-1-\max(\floor*{\frac{\ell - 1}{2}}, 0)$, $m_2=\max(\deg(v),\deg(r))$ and $\lambda = \min(\ell, n)$}
\label{GCD_DIV example}
\begin{tabular}{|c|c|ccccccccc|ccccccccc|c|c|c|c|}
$\ell$ & $\delta$ & $f$                       &                           &                           &                           &                           &                           &                           &                           &                           & $g$                       &                           &                           &                           &                           &                           &                           &                           &   & $m_1$ & $\Lambda$ & $m_2$ & $\lambda$ \\
       &          & 0                         & 1                         & 2                         & 3                         & 4                         & 5                         & 6                         & 7                         & 8                         & 0                         & 1                         & 2                         & 3                         & 4                         & 5                         & 6                         & 7                         & 8 &       &           &       &           \\ \hline
0      & 1        & \cellcolor[HTML]{96FFFB}1 & \cellcolor[HTML]{96FFFB}0 & \cellcolor[HTML]{96FFFB}0 & \cellcolor[HTML]{96FFFB}0 & \cellcolor[HTML]{96FFFB}1 & \cellcolor[HTML]{96FFFB}0 & \cellcolor[HTML]{96FFFB}1 & \cellcolor[HTML]{96FFFB}1 & \cellcolor[HTML]{96FFFB}1 & \cellcolor[HTML]{FFFC9E}1 & \cellcolor[HTML]{FFFC9E}0 & \cellcolor[HTML]{FFFC9E}0 & \cellcolor[HTML]{FFFC9E}1 & \cellcolor[HTML]{FFFC9E}0 & \cellcolor[HTML]{FFFC9E}0 & \cellcolor[HTML]{FFFC9E}0 & \cellcolor[HTML]{FFFC9E}1 & 0 & 7     & 8         & 0     & 0         \\
1      & 0        & \cellcolor[HTML]{FFFC9E}1 & \cellcolor[HTML]{FFFC9E}0 & \cellcolor[HTML]{FFFC9E}0 & \cellcolor[HTML]{FFFC9E}1 & \cellcolor[HTML]{FFFC9E}0 & \cellcolor[HTML]{FFFC9E}0 & \cellcolor[HTML]{FFFC9E}0 & \cellcolor[HTML]{FFFC9E}1 & 0                         & \cellcolor[HTML]{96FFFB}0 & \cellcolor[HTML]{96FFFB}0 & \cellcolor[HTML]{96FFFB}1 & \cellcolor[HTML]{96FFFB}1 & \cellcolor[HTML]{96FFFB}0 & \cellcolor[HTML]{96FFFB}1 & \cellcolor[HTML]{96FFFB}0 & \cellcolor[HTML]{96FFFB}1 & 1 & 7     & 7         & 1     & 1         \\
2      & 1        & \cellcolor[HTML]{FFFC9E}1 & \cellcolor[HTML]{FFFC9E}0 & \cellcolor[HTML]{FFFC9E}0 & \cellcolor[HTML]{FFFC9E}1 & \cellcolor[HTML]{FFFC9E}0 & \cellcolor[HTML]{FFFC9E}0 & \cellcolor[HTML]{FFFC9E}0 & \cellcolor[HTML]{FFFC9E}1 & 0                         & \cellcolor[HTML]{96FFFB}0 & \cellcolor[HTML]{96FFFB}1 & \cellcolor[HTML]{96FFFB}1 & \cellcolor[HTML]{96FFFB}0 & \cellcolor[HTML]{96FFFB}1 & \cellcolor[HTML]{96FFFB}0 & \cellcolor[HTML]{96FFFB}1 & 1                         & 0 & 6     & 7         & 2     & 2         \\
3      & 2        & \cellcolor[HTML]{FFFC9E}1 & \cellcolor[HTML]{FFFC9E}0 & \cellcolor[HTML]{FFFC9E}0 & \cellcolor[HTML]{FFFC9E}1 & \cellcolor[HTML]{FFFC9E}0 & \cellcolor[HTML]{FFFC9E}0 & \cellcolor[HTML]{FFFC9E}0 & \cellcolor[HTML]{FFFC9E}1 & 0                         & \cellcolor[HTML]{96FFFB}1 & \cellcolor[HTML]{96FFFB}1 & \cellcolor[HTML]{96FFFB}0 & \cellcolor[HTML]{96FFFB}1 & \cellcolor[HTML]{96FFFB}0 & \cellcolor[HTML]{96FFFB}1 & 1                         & 0                         & 0 & 5     & 6         & 3     & 3         \\
4      & -1       & \cellcolor[HTML]{96FFFB}1 & \cellcolor[HTML]{96FFFB}1 & \cellcolor[HTML]{96FFFB}0 & \cellcolor[HTML]{96FFFB}1 & \cellcolor[HTML]{96FFFB}0 & \cellcolor[HTML]{96FFFB}1 & 1                         & 0                         & 0                         & \cellcolor[HTML]{FFFC9E}1 & \cellcolor[HTML]{FFFC9E}0 & \cellcolor[HTML]{FFFC9E}0 & \cellcolor[HTML]{FFFC9E}0 & \cellcolor[HTML]{FFFC9E}1 & \cellcolor[HTML]{FFFC9E}0 & \cellcolor[HTML]{FFFC9E}1 & 0                         & 1 & 5     & 6         & 3     & 4         \\
5      & 0        & \cellcolor[HTML]{96FFFB}1 & \cellcolor[HTML]{96FFFB}1 & \cellcolor[HTML]{96FFFB}0 & \cellcolor[HTML]{96FFFB}1 & \cellcolor[HTML]{96FFFB}0 & \cellcolor[HTML]{96FFFB}1 & 1                         & 0                         & 0                         & \cellcolor[HTML]{FFFC9E}1 & \cellcolor[HTML]{FFFC9E}0 & \cellcolor[HTML]{FFFC9E}1 & \cellcolor[HTML]{FFFC9E}1 & \cellcolor[HTML]{FFFC9E}1 & \cellcolor[HTML]{FFFC9E}1 & 0                         & 1                         & 1 & 5     & 5         & 3     & 5         \\
6      & 1        & \cellcolor[HTML]{96FFFB}1 & \cellcolor[HTML]{96FFFB}1 & \cellcolor[HTML]{96FFFB}0 & \cellcolor[HTML]{96FFFB}1 & \cellcolor[HTML]{96FFFB}0 & \cellcolor[HTML]{96FFFB}1 & 1                         & 0                         & 0                         & \cellcolor[HTML]{FFFC9E}1 & \cellcolor[HTML]{FFFC9E}1 & \cellcolor[HTML]{FFFC9E}0 & \cellcolor[HTML]{FFFC9E}1 & \cellcolor[HTML]{FFFC9E}0 & 0                         & 1                         & 1                         & 1 & 3     & 5         & 3     & 6         \\
7      & 0        & \cellcolor[HTML]{FFFC9E}1 & \cellcolor[HTML]{FFFC9E}1 & \cellcolor[HTML]{FFFC9E}0 & \cellcolor[HTML]{FFFC9E}1 & \cellcolor[HTML]{FFFC9E}0 & 0                         & 1                         & 1                         & 1                         & \cellcolor[HTML]{96FFFB}0 & \cellcolor[HTML]{96FFFB}0 & \cellcolor[HTML]{96FFFB}0 & \cellcolor[HTML]{96FFFB}0 & \cellcolor[HTML]{96FFFB}1 & 1                         & 0                         & 0                         & 1 & 3     & 4         & 4     & 7         \\
8      & 1        & \cellcolor[HTML]{FFFC9E}1 & \cellcolor[HTML]{FFFC9E}1 & \cellcolor[HTML]{FFFC9E}0 & \cellcolor[HTML]{FFFC9E}1 & \cellcolor[HTML]{FFFC9E}0 & 0                         & 1                         & 1                         & 1                         & \cellcolor[HTML]{96FFFB}0 & \cellcolor[HTML]{96FFFB}0 & \cellcolor[HTML]{96FFFB}0 & \cellcolor[HTML]{96FFFB}1 & 1                         & 0                         & 0                         & 1                         & 0 & 3     & 4         & 5     & 8         \\
9      & 2        & \cellcolor[HTML]{FFFC9E}1 & \cellcolor[HTML]{FFFC9E}1 & \cellcolor[HTML]{FFFC9E}0 & \cellcolor[HTML]{FFFC9E}1 & \cellcolor[HTML]{FFFC9E}0 & 0                         & 1                         & 1                         & 1                         & \cellcolor[HTML]{96FFFB}0 & \cellcolor[HTML]{96FFFB}0 & \cellcolor[HTML]{96FFFB}1 & 1                         & 0                         & 0                         & 1                         & 0                         & 0 & 2     & 3         & 6     & 8         \\
10     & 3        & \cellcolor[HTML]{FFFC9E}1 & \cellcolor[HTML]{FFFC9E}1 & \cellcolor[HTML]{FFFC9E}0 & \cellcolor[HTML]{FFFC9E}1 & \cellcolor[HTML]{FFFC9E}0 & 0                         & 1                         & 1                         & 1                         & \cellcolor[HTML]{96FFFB}0 & \cellcolor[HTML]{96FFFB}1 & 1                         & 0                         & 0                         & 1                         & 0                         & 0                         & 0 & 1     & 3         & 7     & 8         \\
11     & 4        & \cellcolor[HTML]{FFFC9E}1 & \cellcolor[HTML]{FFFC9E}1 & \cellcolor[HTML]{FFFC9E}0 & \cellcolor[HTML]{FFFC9E}1 & \cellcolor[HTML]{FFFC9E}0 & 0                         & 1                         & 1                         & 1                         & \cellcolor[HTML]{96FFFB}1 & 1                         & 0                         & 0                         & 1                         & 0                         & 0                         & 0                         & 0 & 0     & 2         & 8     & 8         \\
12     & -3       & \cellcolor[HTML]{96FFFB}1 & 1                         & 0                         & 0                         & 1                         & 0                         & 0                         & 0                         & 0                         & \cellcolor[HTML]{FFFC9E}1 & \cellcolor[HTML]{FFFC9E}0 & \cellcolor[HTML]{FFFC9E}1 & \cellcolor[HTML]{FFFC9E}0 & 0                         & 1                         & 1                         & 1                         & 1 & 0     & 2         & 8     & 8         \\
13     & -2       & \cellcolor[HTML]{96FFFB}1 & 1                         & 0                         & 0                         & 1                         & 0                         & 0                         & 0                         & 0                         & \cellcolor[HTML]{FFFC9E}0 & \cellcolor[HTML]{FFFC9E}1 & \cellcolor[HTML]{FFFC9E}0 & 0                         & 1                         & 1                         & 1                         & 1                         & 1 & 0     & 1         & 8     & 8         \\
14     & -1       & \cellcolor[HTML]{96FFFB}1 & 1                         & 0                         & 0                         & 1                         & 0                         & 0                         & 0                         & 0                         & \cellcolor[HTML]{FFFC9E}1 & \cellcolor[HTML]{FFFC9E}0 & 0                         & 1                         & 1                         & 1                         & 1                         & 1                         & 0 & 0     & 1         & 8     & 8         \\
15     & 0        & \cellcolor[HTML]{96FFFB}1 & 1                         & 0                         & 0                         & 1                         & 0                         & 0                         & 0                         & 0                         & \cellcolor[HTML]{FFFC9E}0 & 0                         & 1                         & 1                         & 1                         & 1                         & 1                         & 0                         & 1 & 0     & 0         & 8     & 8        
\end{tabular}
}}
\end{table}
\section{Results}
\label{Results}
Table \ref{Qubit and gate count comparison} shows the comparison of quantum resource needed to compute the division algorithm in binary field. We used Python to calculate the number of qubits and Toffoli gates.

\begin{table}[H]
\caption{Qubit and gate count for the division algorithm in a binary field.}
\label{Qubit and gate count comparison}
\centerline{
\begin{tabular}{|c|cc|cc|}
\hline
    & \multicolumn{2}{c|}{Original GCD\_DIV} & \multicolumn{2}{c|}{New GCD\_DIV} \\
n   & qubits           & TOF gates           & qubits         & TOF gates        \\ \hline
8   & 67               & 2954                & 64             & 3726             \\
16  & 123              & 7594                & 116            & 10190            \\
127 & 907              & 252746              & 839            & 390370           \\
163 & 1155             & 409174              & 1074           & 635366           \\
233 & 1645             & 780734              & 1529           & 1234566          \\
283 & 1995             & 1118134             & 1854           & 1782566          \\
571 & 4012             & 4279890             & 3727           & 6945258          \\ \hline
\end{tabular}
}
\end{table}

If we apply the new GCD\_DIV gate to the circuit of Shor's algorithm in \cite{banegas2020concrete}, the total number of Toffoli gates is roughly $4n$ times of the number of Toffoli gates required in the division gate. On the other hand, the total number of qubits is increased by only one from the number of qubits used in the division algorithm, which makes the whole circuit space-optimized.
\section{Conclusion}
\label{Conclusion}
Table \ref{Qubit and gate count comparison} shows the concrete numbers of logical qubits required in the division algorithm in a binary field $\mathbb{F}_{2^n}$. The new GCD algorithm requires roughly $n/2$ fewer qubits than the GCD algorithm used in \cite{banegas2020concrete}. This space-optimized result is obtained from an observation of the changes of the qubit arrays in the previous GCD algorithm\cite{banegas2020concrete}. The fact that the sizes of $f, g$ constantly decrease while the sizes of $v, r$ constantly increase is the key to our idea. Although the new algorithm requires more quantum gates to manage $mask$ register, this does not consume additional ancillary qubits. So, we are luckily able to reduce the number of qubits.

\bibliographystyle{alpha}
\bibliography{reference}

\appendix
\section{Large Controlled Leftshift Gate}
\label{appendix:Large Controlled Leftshift}
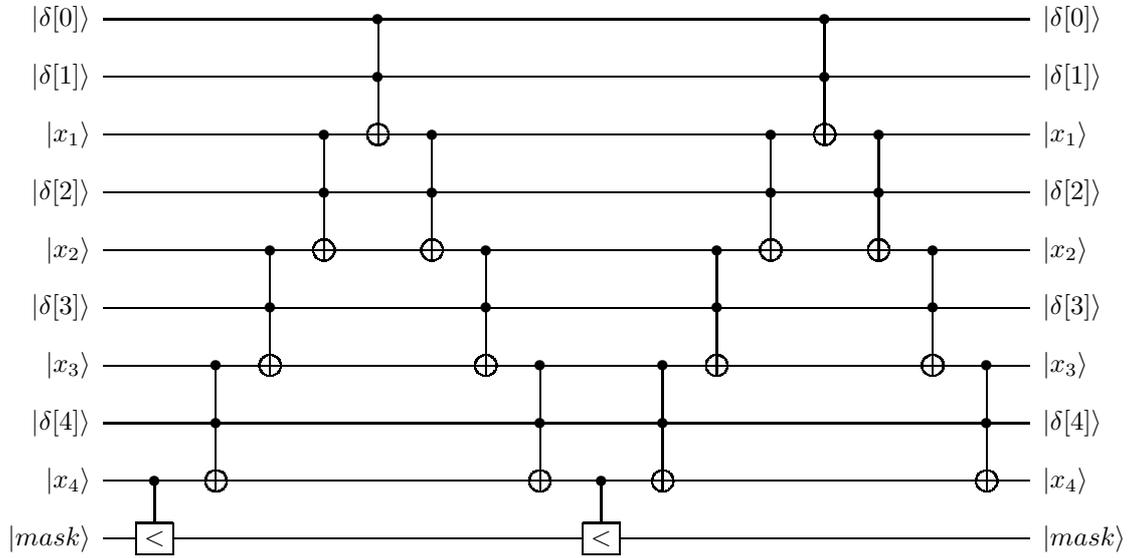
\begin{figure}[H]
\centerline{
\Qcircuit @C=1.2em @R=1em @!R {
\lstick{\ket{\delta[0]}} & \qw      & \qw      & \qw      & \qw      & \ctrl{2} & \qw      & \qw      & \qw      & \qw      & \qw      & \qw      & \qw      & \ctrl{2} & \qw      & \qw      & \qw      & \rstick{\ket{\delta[0]}} \qw \\
\lstick{\ket{\delta[1]}} & \qw      & \qw      & \qw      & \qw      & \ctrl{1} & \qw      & \qw      & \qw      & \qw      & \qw      & \qw      & \qw      & \ctrl{1} & \qw      & \qw      & \qw      & \rstick{\ket{\delta[1]}} \qw \\
\lstick{\ket{x_1}}       & \qw      & \qw      & \qw      & \ctrl{2} & \targ    & \ctrl{2} & \qw      & \qw      & \qw      & \qw      & \qw      & \ctrl{2} & \targ    & \ctrl{2} & \qw      & \qw      & \rstick{\ket{x_1}} \qw \\
\lstick{\ket{\delta[2]}} & \qw      & \qw      & \qw      & \ctrl{1} & \qw      & \ctrl{1} & \qw      & \qw      & \qw      & \qw      & \qw      & \ctrl{1} & \qw      & \ctrl{1} & \qw      & \qw      & \rstick{\ket{\delta[2]}} \qw \\
\lstick{\ket{x_2}}       & \qw      & \qw      & \ctrl{2} & \targ    & \qw      & \targ    & \ctrl{2} & \qw      & \qw      & \qw      & \ctrl{2} & \targ    & \qw      & \targ    & \ctrl{2} & \qw      & \rstick{\ket{x_2}} \qw \\
\lstick{\ket{\delta[3]}} & \qw      & \qw      & \ctrl{1} & \qw      & \qw      & \qw      & \ctrl{1} & \qw      & \qw      & \qw      & \ctrl{1} & \qw      & \qw      & \qw      & \ctrl{1} & \qw      & \rstick{\ket{\delta[3]}} \qw \\
\lstick{\ket{x_3}}       & \qw      & \ctrl{2} & \targ    & \qw      & \qw      & \qw      & \targ    & \ctrl{2} & \qw      & \ctrl{2} & \targ    & \qw      & \qw      & \qw      & \targ    & \ctrl{2} & \rstick{\ket{x_3}} \qw \\
\lstick{\ket{\delta[4]}} & \qw      & \ctrl{1} & \qw      & \qw      & \qw      & \qw      & \qw      & \ctrl{1} & \qw      & \ctrl{1} & \qw      & \qw      & \qw      & \qw      & \qw      & \ctrl{1} & \rstick{\ket{\delta[4]}} \qw \\
\lstick{\ket{x_4}}       & \ctrl{1} & \targ    & \qw      & \qw      & \qw      & \qw      & \qw      & \targ    & \ctrl{1} & \targ    & \qw      & \qw      & \qw      & \qw      & \qw      & \targ    & \rstick{\ket{x_4}} \qw \\
\lstick{\ket{mask}}      & \gate{<} & \qw      & \qw      & \qw      & \qw      & \qw      & \qw      & \qw      & \gate{<} & \qw      & \qw      & \qw      & \qw      & \qw      & \qw      & \qw      & \rstick{\ket{mask}} \qw \\
}
}
\caption{The quantum circuit for $C^n\mathrm{MASK\_LEFTSHIFT}_{\delta}(\cdot)$ gate where $n=5$ and $x_1, x_2, x_3, x_4$ are borrowed bits.}
\label{LargeControlledLeftshiftCircuit}
\end{figure}

We reference the "Constructing a large controlled nots" in \cite{gidney2015constructing} to construct large controlled shift gate $\mathrm{C^nMASK\_LEFTSHIFT}_{\delta}(\cdot)$ used in the algorithm \ref{NewGCDAlgorithm}. The quantum circuit is designed as figure \ref{LargeControlledLeftshiftCircuit}. When the size of a control register $\delta$ is $l$ and the input size is $m$, $4l+m-7$ Toffoli gates, $2m-2$ CNOT gates and $l-1$ borrowed bits are needed. In the case of the algorithm \ref{NewGCDAlgorithm}, $l=\floor*{\log(n)} + 1$ and $m = \Lambda + 1$. So, $4\floor*{\log(n)} + \Lambda -2$ Toffoli gates, $2\Lambda$ CNOT gates and $\floor*{\log(n)}$ borrowed bits are needed. In this circuit, we use algorithm $\ref{ControlledLeftshiftAlgorithm}$ for the LEFTSHIFT gate for $mask$ register.

\end{document}